\documentstyle[twocolumn,aps]{revtex} 
 
\draft 
\begin{document} 
 
 
\title{Melting and Dimensionality of the Vortex Lattice in Underdoped 
YBa$_2$Cu$_3$O$_{6.60}$} 
 
\author{ J.~E.~Sonier$^1$, J.~H.~Brewer$^2$, R.~F.~Kiefl$^2$, 
D~.A.~Bonn$^2$, J.~Chakhalian$^2$, S.~R.~Dunsiger$^2$, 
W.~N.~Hardy$^2$, R.~Liang$^2$, W.A.~MacFarlane$^3$,
R.~I.~Miller$^2$, D.~R.~Noakes$^4$, T.~M.~Riseman$^5$ and C.~E.~Stronach$^4$}
\address{$^1$Los Alamos National Laboratory, Los Alamos, New Mexico, USA 87545}
\address{$^2$TRIUMF, Canadian Institute for Advanced Research 
and Department of Physics and Astronomy, University of British Columbia, 
Vancouver, British Columbia, Canada V6T 1Z1} 
\address{$^3$Laboratoire de Physique des Solides, UMR 8502,
Universit\'{e} de Paris-Sud, 91405 Orsay Cedex, France}
\address{$^4$Department of Physics, Virginia State University, 
Petersburg, Virginia 23806}
\address{$^5$Superconductivity Research Group, University of Birmingham,
Birmingham, B15 2TT, United Kingdom}

\date{June 29, 1999} 
\date{ \rule{2.5in}{0pt} } 
 
\maketitle 
\begin{abstract} \noindent 
Muon spin rotation ($\mu$SR) measurements of the
magnetic field distribution in the vortex state of the oxygen 
deficient high-$T_c$ superconductor YBa$_2$Cu$_3$O$_{6.60}$
reveal a vortex-lattice melting transition at much lower temperature
than that in the fully oxygenated material.
The transition is best described by a model in which adjacent layers
of ``pancake'' vortices decouple in the liquid phase. Evidence is
also found for a pinning-induced crossover from a solid 3D to 
quasi-2D vortex lattice, similar to that observed 
in the highly anisotropic
superconductor Bi$_{2+x}$Sr$_{2-x}$CaCu$_2$O$_{8+y}$.    
\end{abstract} 

In the high-temperature superconductors there exist several
exotic vortex lattice (VL) phases owing to the weak coupling between
the superconducting CuO$_2$ layers (which gives rise to highly flexible vortices),
a relatively short coherence length (which enhances
their susceptibility to pinning) and high values of $T_c$ (which
permit large thermal energies to be reached in the vortex state). 
To date much attention has been paid to VL melting and pinning
effects in the highly anisotropic
superconductor Bi$_{2+x}$Sr$_{2-x}$CaCu$_2$O$_{8+y}$ (BSCCO)
and in YBa$_2$Cu$_3$O$_{7-\delta}$ (YBCO) for near optimal
oxygen concentrations
({\it i.e.} $\delta \! \approx \! 0.05$, giving the highest value
of $T_c$). The vortices in BSCCO
are often described as stacks of two-dimensional (2D) ``pancakes'', 
which can become misaligned at elevated temperatures and/or in the 
presence of strong random pinning sites. On the other hand, in 
``optimally'' doped YBCO the vortices behave
in a three-dimensional (3D) manner. This difference is attributed to
the mass anisotropy $\gamma \! = \! (m_c/ m_{ab})^{1/2} \! 
= \! \lambda_c / \lambda_{ab} \! \approx \! 5$-$7$ in YBCO and 
$\gamma \! \approx \! 50$-$250$ in BSCCO \cite{Hussey:96},
where $\lambda_{ab}$ and $\lambda_c$ are the magnetic penetration
depths describing the screening of flux by supercurrents flowing
in and out of the CuO$_2$ layers, respectively. 
An added complication in many studies of YBCO is the 
presence of twin-plane boundaries which may act as extended 
pinning sites for vortices.

The muon spin rotation ($\mu$SR) technique measures the
muon precession frequency distribution and thus the internal
magnetic field distribution $n(B)$, also known as the $\mu$SR
line shape, which has proven to be a powerful tool for investigating 
VL phases \cite{Lee:93,Riseman:95,Sonier:97}. 
In optimally doped YBCO, strong local pinning broadens 
the $\mu$SR line shape \cite{Harshman:91} as predicted
for small random displacements of 3D vortex lines \cite{Brandt:91}.     
By contrast, random pinning in BSCCO leads to a field-induced 
reduction of the $\mu$SR line width 
\cite{Lee:93,Harshman:93,Bernhard:95,Aegerter:96}
associated with a dimensional crossover from 
a 3D-VL to a quasi-2D system consisting of independent
weakly interacting VLs in different CuO$_2$ layers. 
In the quasi-2D region, the repulsive interaction between pancake 
vortices in the layers exceeds the strength of their interlayer coupling, 
so that random pinning results in a   
misalignment of the pancake vortices in the field direction.

The effect on the $\mu$SR line shape of thermal fluctuations 
of the vortex positions has been the focus of numerous studies
in BSCCO \cite{Lee:93,Lee:95,Lee:97,Blasius:99}.  
As explained in Ref.~\cite{Lee:95}, the typical time scale for 
vortex fluctuations is short enough that the
muon detects a fluctuation-averaged field. This results in
a reduction of both the width and the {\em skewness} of the 
$\mu$SR line shape. In clean samples the VL melts in a 
first-order phase transition (see, for example, Ref.~\cite{Safar:92}).  
Recent $\mu$SR measurements in BSCCO have been
interpreted as evidence for a two-stage VL transition: first
the {\em intralayer} coupling of the pancake vortices is
overcome by thermal fluctuations, then their {\em interlayer}
coupling is lost \cite{Blasius:99}.      

In the less anisotropic compound YBCO, a melting transition
has been observed by small-angle neutron scattering,
magnetization, transport, specific heat and Hall probe ac susceptibility
measurements (see, for example, Ref~\cite{YBCOmelt}). 
However, unlike in BSCCO, the vortex-liquid phase is found
only in a very narrow region below the
second-order phase transition at $B_{c2} (T)$. A melting transition near
$B_{c2} (T)$ is difficult to study with the $\mu$SR technique, because 
even for a 3D VL, the $\mu$SR line shape narrows and 
becomes more symmetric due to the overlap of the vortices.
  
It is well established that removal of oxygen from
the CuO chain layers in YBCO weakens the coupling between
the superconducting CuO$_2$ layers. Here we report 
$\mu$SR measurements of the VL
in the oxygen-deficient compound YBa$_2$Cu$_3$O$_{6.60}$,
which has an increased anisotropy ratio 
$\gamma \! \approx \! 22$-$36$ (see, Ref.~\cite{Janossy:91}).
We observe an expanded vortex-liquid region and a pinning-induced 
crossover from a 3D to a quasi-2D system, qualitatively resembling the
vortex phase diagram of BSCCO. 
  
The $\mu$SR experiments were performed on the M15 and M20 surface
muon beamlines at TRIUMF, Canada, using the experimental setup
described in Ref.~\cite{Sonier:94}. Measurements of the internal
magnetic field distribution
were made in both {\em twinned} and {\em detwinned}
crystals of underdoped YBa$_2$Cu$_3$O$_{6.60}$, with
superconducting transition temperatures $T_c \! = \! 59(0.1)$~K. These
samples were previously studied with $\mu$SR at low
temperatures \cite{Sonier:97}. The crystals were mounted with their
$\hat{c}$-axis parallel to both the applied magnetic field 
\boldmath$H$\unboldmath and the
muon beam direction. The positive muons were injected into the sample
with their initial spin polarization
perpendicular to \boldmath$H$\unboldmath. 
As described fully elsewhere \cite{Brewer:94}, 
$n(B)$ or the muon precession frequency distribution
$n(\omega \! = \! \gamma_{\mu} B)$, where $\gamma_{\mu}$ is the
muon gyromagnetic ratio, is obtained by monitoring the $e^{+}$
count rate as the muon decay pattern sweeps by the positron
detectors.     
 
Figure~1 shows the evolution of the fast Fourier transform 
(FFT) of the muon precession signal upon warming the
sample, following field cooling to $T \! = \! 2.4$~K in
a magnetic field $H \! = \! 1.49$~T. 
At $T \! = \! 20.5$~K the FFT shows the basic features 
expected for a well-ordered 3D solid VL --- in particular 
a high-field ``tail'' associated with the region close to 
the vortex cores. 
However, at $T \! = \! 40$~K the FFT is completely symmetric and the
line width is drastically reduced --- both of which
characteristize a vortex-liquid phase. Despite these obvious 
changes in the $\mu$SR line shape, the melting transition cannot 
usually be determined accurately by visual inspection, because
the output frequency spectrum is artificially
broadened by the finite time range and the ``apodization'' needed
to eliminate ``ringing'' in the FFT \cite{Fleming:82}. 
To quantify these changes in the field distribution we calculate
the {\it skewness} parameter \cite{Lee:93},
$\alpha \! = \! \langle (\delta B)^3 \rangle^{1/3} /
\langle (\delta B)^2 \rangle^{1/2}$, where
$\langle (\delta B)^n \rangle \! = \! 
\langle (B \! - \! \langle B \rangle)^n \rangle$. 
To obtain reliable values for the moments 
$\langle (\delta B)^n \rangle$ we fit the $\mu$SR spectra
in the time domain to a polarization function calculated
assuming a Ginzburg-Landau (GL) model for the VL field 
profile $B({\bf r})$, as described in Ref.~\cite{Sonier:97}. 
The fitted function $B({\bf r})$ properly
accounts for the long high-field ``tail'' in $n(B)$ when the
VL is 3D. This method of analysis also provides a means of 
monitoring the $\mu$SR line width through the fitted value of 
$1/\lambda_{ab}^2$. A hexagonal arrangement 
of vortices was assumed, consistent with recent neutron
scattering measurements on {\em untwinned} YBCO \cite{Johnson:99}.  
   
Figure~2 shows the temperature dependence of $\alpha$ and 
$1/\lambda_{ab}^2$ at three of the fields considered. The observed 
values of $\alpha \! \approx \! 1.2$ are in agreement with
values previously obtained in the 3D-VL phase of
BSCCO \cite{Lee:93,Bernhard:95,Aegerter:96,Lee:97,Blasius:99}
and that predicted for $s$- and $d$-wave VLs 
at low reduced fields $B/B_{c2}$ \cite{Ichioka:99}.
For temperatures above $T_{\rm m}$ the fitted values 
of $1/\lambda_{ab}^2$ and hence $\alpha$ drop to zero. 
This does not mean that the superfluid density has decreased to zero, 
but rather that the $\mu$SR spectrum no longer contains a component 
corresponding to an ordered 3D vortex structure. For 
$T \! > \! T_{\rm m}$ the $\mu$SR spectra fit well to 
a single Gaussian function, with a line width much greater
than that due to nuclear dipoles.

In Ref.~\cite{Rodriguez:98} a sharp transition was observed
in ac susceptibility measurements on twinned
YBa$_2$Cu$_3$O$_{6.60}$ crystals below a characteristic
field $H^* \! \approx \! 0.07$~T, possibly 
associated with a melting transition. Above $H^*$, where the
transition was found to be continuous, it was suggested that a quasi-2D
system exists which is highly sensitive to pinning-induced disorder.
In the present study, where $H \! > \! H^*$, the field distributions
for $T \! < \! T_{\rm m}$ and $0.1 \! < \!  H \! < \! 1.49$~T fit well
to the GL model for an ordered 3D VL. 
The rms deviation of the vortices
from their ideal positions in the hexagonal VL 
was found \cite{Sonier:97} to be $\! < \! 8$~\%. 
These results suggest that the VL for our samples 
in this region of the $B$-$T$ phase diagram
exhibits a 3D behavior, most likely
consisting of stacks of strongly coupled pancake vortices.
Above $T_{\rm m}$, thermal fluctuations of the vortices lead to a
loss of long-range spatial order.

The top panel of Figure~3 shows the $\mu$SR line shape in
the {\em twinned} sample of YBa$_2$Cu$_3$O$_{6.60}$ after
cooling in a field $H \! \approx \! 2.89$~T to $T \! = \! 2.5$~K, 
followed by an increase of $ \Delta H \approx \! 0.01$~T.
Notice that the small background signal is positioned at the external
field $H \! = \! 2.90$~T, whereas the signal originating from
the sample looks as though the external field is 
still $H \! = \! 2.89$~T.
This implies that the VL is strongly pinned. However, unlike in the
line shapes observed at the lower fields, the ``tail'' appears on
the {\em low}-field side of the cusp. Numerical calculations performed in 
Ref.~\cite{Schneider:95}, which account for the
sample geometry effect, show that such a line shape can originate
from a system of 2D pancake vortices that are ordered in the
planes but uncorrelated between adjacent layers. 
The low-field ``tail'' is specifically associated with 
a lower flux density at the sample edges due to 
a nonuniform demagnetization. Thus clear evidence is
observed for a pinning-induced dimensional crossover above a critical
field $B_{\rm cr}(T)$.

By contrast, at the same field $H \! = \! 2.9$~T in the
{\em detwinned} sample the ``tail'' extends to the {\em high}-field 
side of the distribution, as expected
for an ordered 3D vortex structure. 
It is tempting to attribute the opposite skewness of the
$\mu$SR line shapes in the two samples to the presence 
of twin planes. However, because
the twin planes extend the full depth of the sample, they would
displace the 2D VLs by the same amount in all layers. 
Since the two samples
were not from the same growth batch, the difference is more likely
related to the concentration and randomness of pointlike defects.
This result stresses the sensitivity of the VL structure to disorder
at this field. 

The middle panel of Fig.~3 shows that the vortices begin
to depin at $T_{\rm p} \! \approx \! 19.8$~K due to thermal fluctuations, and
reposition themselves with an average field $B \! \approx \! 2.90$~T
at $T \! = \! 23$~K (bottom panel of Fig.~3). Additional $\mu$SR spectra
taken without shifting the field show that the VL melts at
$T_{\rm m} \! \approx \! 20$~K.

Figure~4 summarizes our results in a phase diagram. The upper
critical field line represents an approximation
assuming $B_{c2}(T) \! = \! B_c(0)[1 \! - \! (T/T_c)^2]$ and
$B_{c2}(0) \! = \! 70$~T. The data for the twinned and detwinned
samples appear to fall on the same melting line. A fit of the
data below $H \! = \! 2.9$~T 
to the phenomenological melting curve \cite{Lee:97}  
\begin{equation}
B_{\rm m} (T) = \frac{K}{\lambda_{ab}^m (T)T} \, ,
\label{eq:melt}
\end{equation}
yields $m \! = \! 1.6(1)$, where 
$\lambda_{ab} \! = \! \lambda_{ab}(0) [1-(T/T_c)^4]^{-1/2}$,
and $K$ and $m$ are constants. This is well below either
the predicted value $m \! = \! 4$ when interlayer coupling 
of the pancake vortices
is dominated by electromagnetic interactions,
or $m \! = \! 3$ when Josephson coupling 
becomes important \cite{Blatter:96}.
As discussed in Refs.~\cite{Lee:97,Blatter:96}, for finite
Josephson coupling an exponent $m \! = \! 2$ is expected for 
an additional thermodynamic transition at $B \! < \! B_{\rm cr}$
which decouples the layers in the liquid phase.
The decoupling line is predicted to be \cite{Blatter:96}
\begin{equation}
B^{\rm J}_{\rm dc}(T) = \frac{\Phi_0^3 c_D}{4 \pi \mu_0 k_B s \gamma^2}
\frac{1}{\lambda_{ab}^2}{T} \, ,
\label{eq:dc}
\end{equation}
where $c_D \! \sim \! 0.1$ is a decoupling constant and $s$ is the
interlayer spacing. The dashed curve in Fig.~4 is a fit to
Eq.~(\ref{eq:dc}). Taking $s \! = \! 11.7$~\AA 
\cite{Rodriguez:98}, 
$\lambda_{ab}(0) \! = \! 1642$~\AA \cite{Sonier:97} and
$\gamma \! = \! 22$-$36$ \cite{Janossy:91},
we calculate $c_D \! = \! 0.025$-$0.067$ from the fit.
A value $c_D \! = \! 0.076$ was obtained for BSCCO 
in Ref.~\cite{Lee:97}. 
To determine whether the melting transition ({\em i.e.}
to a liquid of vortex lines) coincides 
with the low-field interlayer decoupling transition, 
more measurements in the vicinity of the 
phase transition are needed. We note
that, contrary to the conclusion in Ref.~\cite{Lee:97}, recent
detailed $\mu$SR measurements in BSCCO \cite{Blasius:99}
suggest that $B_{\rm m}(T) \! < \! B^{\rm J}_{\rm dc} (T)$. 

In conclusion, we have observed changes in the $\mu$SR 
line shape of underdoped YBa$_2$Cu$_3$O$_{6.60}$ which
identify a VL melting transition far below that
of optimally doped YBCO. More precisely, fits to the data
suggest that there is a decoupling transition 
from a liquid of vortex lines to a liquid of 2D vortices. 
Our measurements also establish
the existence of a pinning-induced dimensionl crossover
to a quasi-2D vortex system, similar to that observed 
in the highly anisotropic material BSCCO.

The work at TRIUMF was supported by NSERC Canada and 
US AFOSR Grant No. F49620-97-1-0297. The work at Los
Alamos was performed under the auspicies of the US
Department of Energy.
    
\newpage 
\begin{center} 
FIGURE CAPTIONS 
\end{center} 
 
Figure~1. Fourier transform of the muon precession signal in 
twinned YBa$_2$Cu$_3$O$_{6.60}$ at
$T \! = \! 20.5$~K and $T \! = \! 40$~K, in a field
$H \! = \! 1.49$~T.\\

Figure~2. Temperature dependence of the {\em skewness} $\alpha$
[top] and $1/ \lambda_{ab}^2$ [bottom] in 
detwinned YBa$_2$Cu$_3$O$_{6.60}$ at $H \! = \! 1.25$~T [crosses]
and in twinned YBa$_2$Cu$_3$O$_{6.60}$
at applied magnetic fields $H \! = \! 0.74$ [open circles]
and $1.49$~T [solid circles].\\
 
Figure~3. Fourier transforms of muon precession signals in twinned
YBa$_2$Cu$_3$O$_{6.60}$. Top panel: after field cooling at 
$H \! \approx \! 2.89$~T to $T \! = \! 2.5$~K followed by an 
increase in the field to 2.90~T. Middle and bottom panels:
after subsequently warming the sample to $T \! = \! 19.8$~K 
and $23.0$~K, respectively.\\ 

Figure~4. The vortex $B$-$T$ phase diagram
for YBa$_2$Cu$_3$O$_{6.60}$. Open and solid circles
correspond to the twinnned and detwinned samples, respectively.
The dashed curve $B^{\rm J}_{\rm dc}(T)$ is a fit to Eq.~(\ref{eq:dc}),
the solid curve $B_{c2}(T)$ is a theoretical line for the
upper critical field (see text)
and the solid curve $B_{\rm m}(T)$ is a fit of all of the data
to Eq.~(\ref{eq:melt}). The transition line $B_{\rm cr} (T)$ is
approximate.\\

 
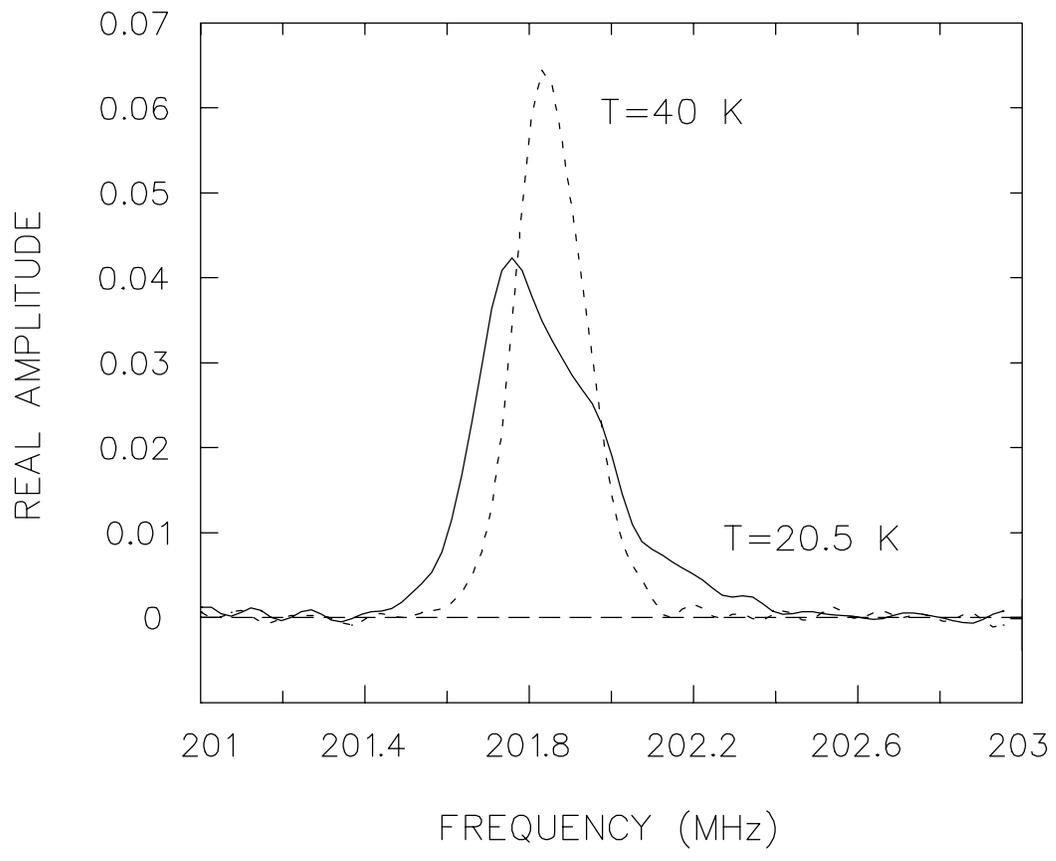
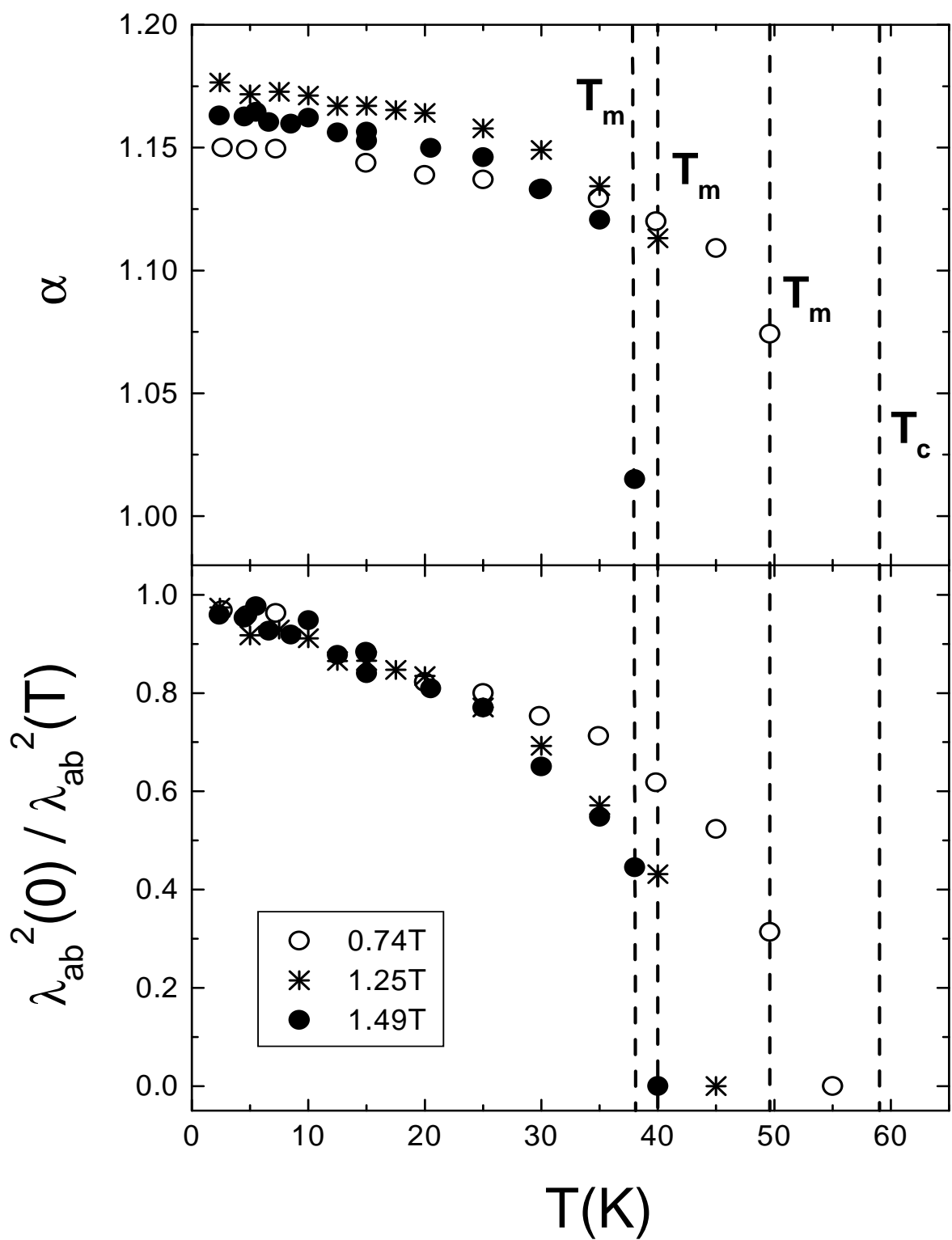
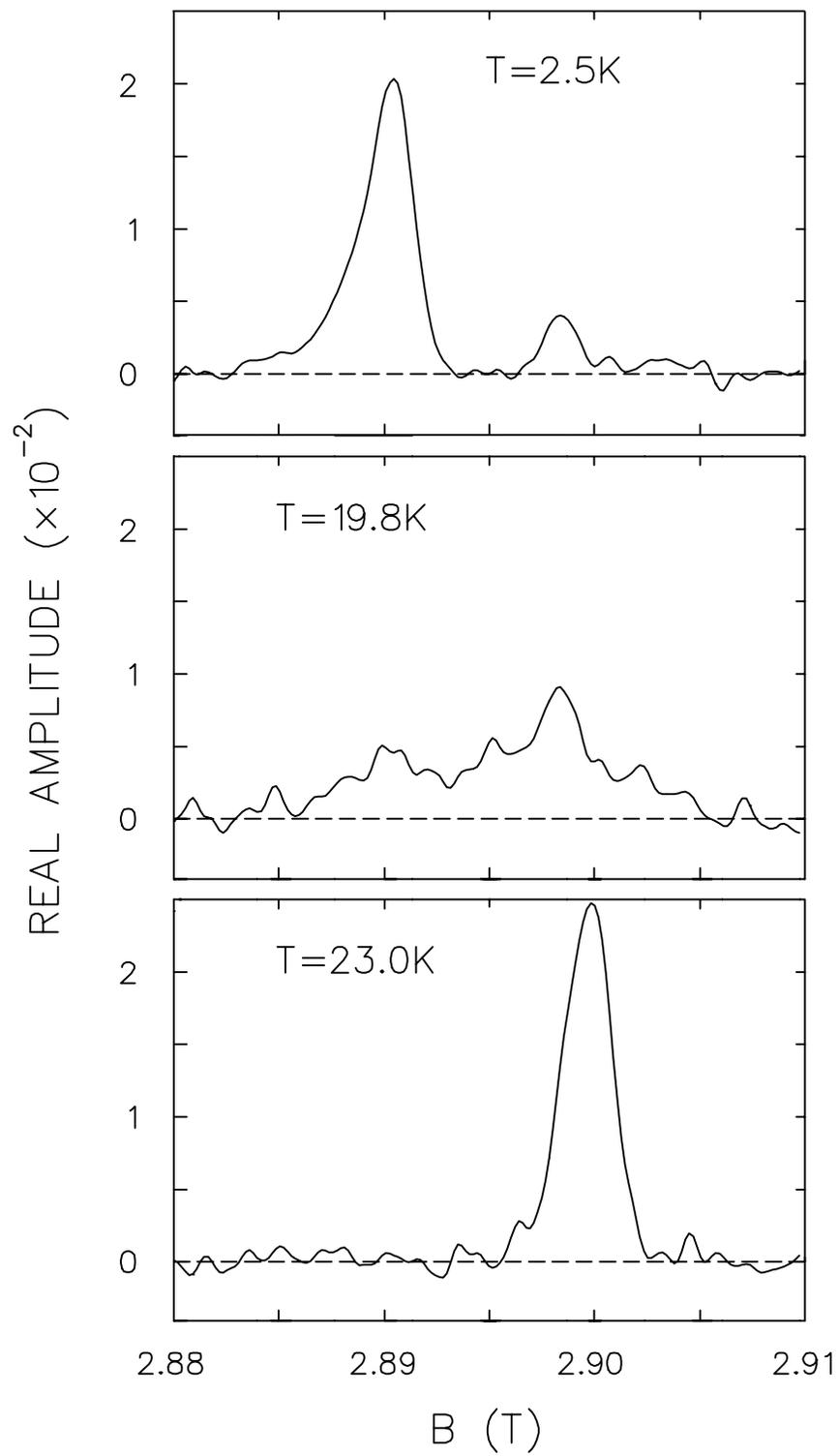
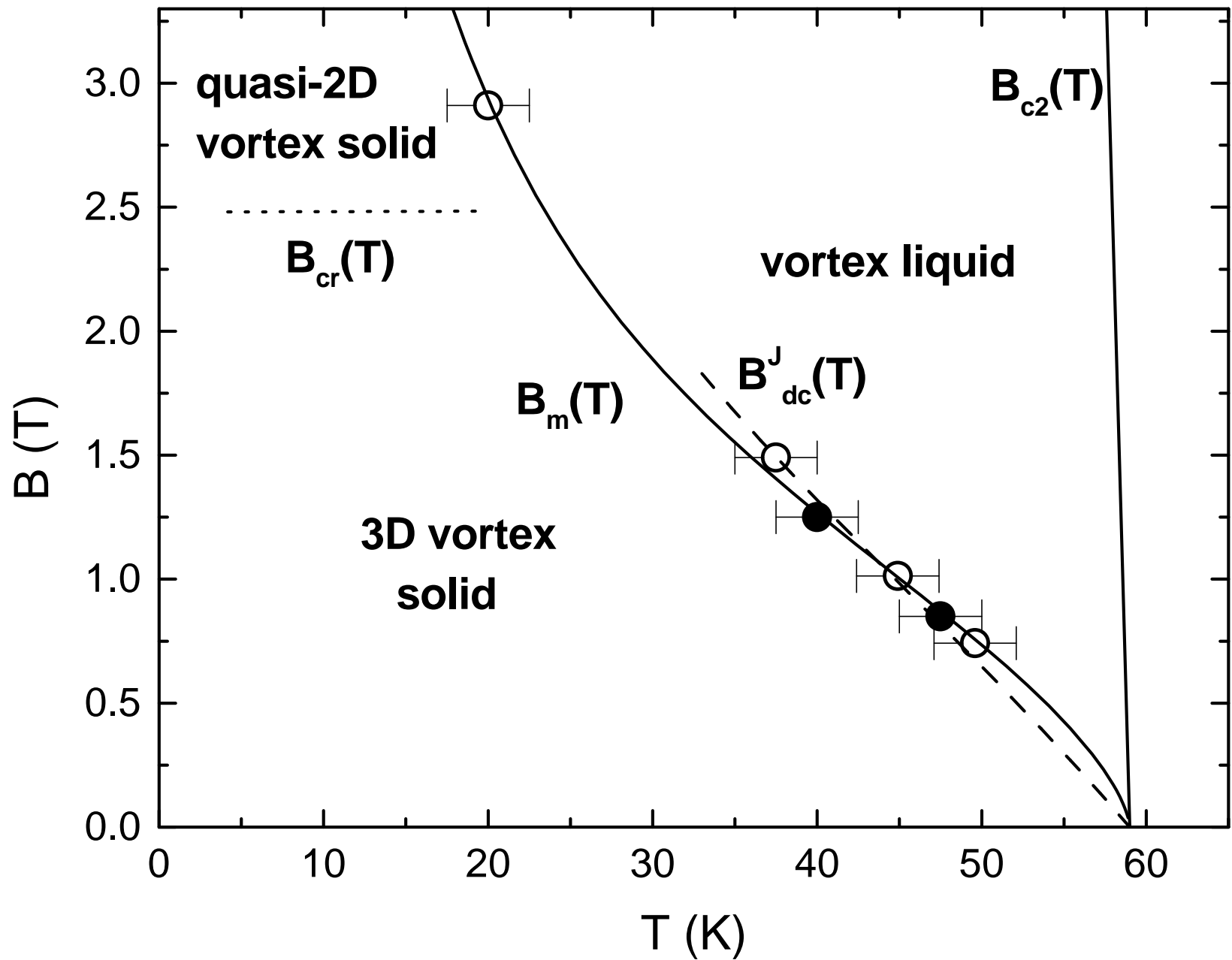
\end{document}